\documentstyle[preprint,aps]{revtex}
 
\def\be{\begin{equation}}
\def\ee{\end{equation}}
\newcommand{\bea}{\begin{eqnarray}}
\newcommand{\eea}{\end{eqnarray}}
\def\we{\approx}
\def\rla{\leftrightarrow}
\def\fr{\frac}
\def\a{\alpha}
\def\b{\beta}

\def\d{\delta}
\def\e{\epsilon}

\def\l{\lambda}
\def\m{\mu}
\def\n{\nu}
\def\r{\rho}
\def\s{\sigma}

\def\th{\theta}
\def\w{\omega}
\def\W{\Omega}

\def\d{\delta}

\def\L{\Lambda}
\def\CL{\cal L}

\def\p{\partial}
\def\t{\tilde}
\def\T{\Theta}
\def\nn{\noindent}
\begin{document}
\draft
\title{{On the equivalence between topologically and non-topologically massive
abelian gauge theories}}
\author{ E. Harikumar  and M. Sivakumar$^{**}$}
\address{ School of Physics,
 University of Hyderabad\\
 Hyderabad, Andhra Pradesh\\
 500 046  INDIA.}
\footnotetext {$^{**}$ Electronic address: mssp@uohyd.ernet.in }  
\maketitle
\begin{abstract} 

We analyse the equivalence between topologically massive gauge
theory (TMGT) and different formulations of non-topologically massive gauge
theories (NTMGTs) in the canonical approach. The different NTMGTs
studied are St\"uckelberg formulation of (A) a first order formulation 
involving one and two form fields, (B) Proca theory, 
and (C) massive Kalb-Ramond theory. We first quantise these reducible gauge 
systems by using the phase space extension procedure and using it,
identify the phase space variables of NTMGTs
which are equivalent to the canonical variables of TMGT
and show that under this the Hamiltonian also get mapped.
Interestingly it is found that the different NTMGTs are equivalent to
different formulations of TMGTs which differ
only by a total divergence term. We also provide covariant mappings
between the fields in TMGT to NTMGTs at the
level of correlation function.

\end{abstract}
\draft
\newpage

Massive gauge theories have been a subject of intense study with popular
approach being the Higgs mechanism. The fact that the Higgs Boson is yet
experimentally elusive, forces an intense study of alternate models
displaying gauge invariant massive spin-1 fields. One such model which is
being studied at both abelian and non-abelian level \cite{bf} is
toplogically massive gauge theory (TMGT),
known as $B{\wedge}F$ theory, where one and two form gauge fields
are coupled in a gauge invariant way. $B{\wedge}F$ theory has also found
applications in areas like Josphson junction arreys, black hole physics etc,.
\cite{oth}. There also exist other
non-topologically massive gauge theories (NTMGTs), wherein 
St\"uckelberg fields are added with a compensating transformation \cite{st},
 such
that the mass term in the Lagrangian is gauge invariant. The reason for
this nomenclature is, as explained below, the equations of motion for the
former case vanishes as an identity whereas for the latter they  vanish as 
a consequence of other equations of motion. The purpose of
this paper is to study the relationship between these two apparently
different formulations. Equivalence of degrees of freedom between TMGT and
Proca theory and their higher dimensional generalizations has been shown in
\cite{ref}. The NTMGTs studied
 here are St\"uckelberg formulations of
(A) a first order formulation involving one and two form fields, (B) Proca 
theory, and (C) massive Kalb-Ramond(K-R) theory described below  
( cf: (29),(45) and (50) respectively). Non-abelian version of (A),
with out the St\"uckelberg fields have been discussed in \cite{ft}.
The massive St\"uckelberg form of K-R theory has been studied in \cite{at}
and has also occured in a description of confining string \cite{poly}. 
The equivalence
of the first one of the above 
to $B{\wedge}F$ theory has been established recently by phase space path
integral approach \cite{us1}. Duality
equivalence, following Buscher's procedure \cite{bus} was established between 
TMGT and NTMGTs (B and C) \cite{us}. But a complete analysis of the 
equivalence between 
TMGT and NTMGTs from the Hamiltonian formulation is still lacking. Also
these systems by themselves are interesting due to their interesting
constraint structure. In this paper we analyse, from the
canonical frame work, the equivalence between $B{\wedge}F$ theory 
and NTMGTs by  providing  a mapping between the phase space variables 
of  TMGT to free NTMGTs which 
render the same algebra in the respective reduced phase spaces.  This mapping 
between the canonical
variables also map the corresponding canonical Hamiltonians and the
equations of motion. Interestingly it is found that different NTMGTs 
gets mapped to different 
formulations of TMGT which differ only by a total divergence. We point out
that evaluation of Dirac brackets for these reducible systems following the
phase space extension approach developed recently in \cite{rb} by itself is
a new result. We also provide an equivalence at the level of corelators
for the field in configuration space between the TMGT and NTMGTs. This paper is organised in the
following way: In section I we 
study the Hamiltonian formulation of the model given in (1), and quantise
it by enlarging the phase space by introducing a pair of new canonical 
coordinates and obtain the Dirac brackets.
In section II we quantize the three different spin-one NTMGTs,
taking into account of the reducibility in the same way.
In section III, we find the mapping between
the phase space variables of $B{\wedge}F$ theory to the St\"uckelberg 
formulations which generates the same algebra of canonical variables
in the reduced phase space. We conclude with discussion in section IV.
 
\nn We use the metric $g_{\m\n} = {\rm diag}(1, -1 -1 -1)$ and $\e_{0123} = 1$.
 
\noindent{\bf Section I}
 
 The Lagrangian for the abelian $B^{\wedge}F$ theory is given by  
\be
{\CL} ={- \frac{1}{4}} F_{\m\n}F^{\m\n}~+~{\frac{1}{{2 \times 3!}}}
 H_{\m\n\l}H^{\m\n\l} ~+~ {\frac{m}{4}} \e_{\m\n\r\l}B^{\m\n}F^{\r\l},
\ee
\nn where $F_{\m\n}$ and $H_{\m\n\l}$ are the field strengths associated
with the fields $A_\m$ and $B_{\m\n}$.
 The equation of motion following from this Lagrangian are
\be
\p_\m F^{\m\n}~+~ \frac{m}{3!} \e^{\n\a\b\l} H_{\a\b\l} = 0,
\ee
and
\be
 \p_\m H^{\m\n\l} - m~\e^{\n\l\a\b} \p_\a A_\b = 0.
\ee
It is easy to see that this system describes a free spin-one theory by
solving the linearly coupled equations (2) and (3). Since the divergence of (2)
and (3) vanishes as an identity, the Lagrangian (1) is said to describe
TMGT. This should be contrasted with NTMGTs given in section II, where the
divergence of equations of motion vanish only after using other equations
of motion.

 The primary constraint following from the Lagrangian are
$ \Pi_0 ~{\rm and}~\Pi_{0 i}$ which are the momentum conjugates of $A^0 {\rm
and}~ B^{0i}$ respectively. The momenta, from which velocities are 
\nn inverted , are given by,
 \bea
\Pi_i = -F_{0 i}~+~\frac{m}{2} \e_{0 i j k} B^{j k},\\
\Pi_{ij}=~H_{0 i j}.
\eea

The canonical Hamiltonian following from  this Lagrangian is,
 
\bea
{\cal H}_c=-\frac{1}{2} \Pi_i \Pi^i+\frac{1}{4} \Pi_{i j}\Pi^{i j}+
\fr{1}{4} F_{i j}F^{i j} -\fr{1}{2 \times 3!} H_{i j k}H^{i j k}\nonumber\\
+\fr{m^2}{4}B_{l m}B^{l m}+ \fr{m}{2} \e_{0 i j k}\Pi^i B^{j k}
\eea
  
 The consistancy of the primary constraints leads to the secondary constraints,
 \bea
\W~=~ \p_i \Pi^i,\\
\L_i~=~ -\p^j \Pi_{i j}~+~\fr{m}{2}\e_{0 i j k}F^{j k}
\eea
 Note that these are first class and (8) is reducible;
%% that  is to say, all the first class constraints are not linearly 
%%independet. Here the constraints $\L_i$ are not all linearly independet as
$\p^i \L_i =0$.
 The gauge -fixing conditions for these constraint are given by
 \bea
\p^i A_i\we 0,\\
\w^i=\p_j B^{i j}\we 0. 
 \eea
 Here the gauge-fixing condition $\p_j B^{i j}\we 0$ is also not linearly 
 independent.

In usual Dirac procedure, one first isolate the linearly independent 
constraints, and then procede to construct the Dirac brackets \cite{rk}. 
But this elimination can not be done uniquely. More over, in general,
the linearly independent constraints obtained in this way do not generate
all those transformations under which the Lagrangian is invariant. 
Both these deffects can be avoided in the phase
space extension method developed in \cite{rb}. Here one handles the 
reduciblity by
modifying the constraints itself. In this method
one enlarges the phase space by introducing a pair of 
canonically congugate auxiliary variables $p,~{\rm and}~ q$ with
\be
\left\{p(x)~,~ q(y)\right\}_{PB}~=~-\delta({\vec x}-{\vec y}).
\ee
The modified constraint and the coresponding gauge-fixing condition
become,

%..The constraints are modified to include a canonically congugate
%pair$p, q$. The reducible constraints now become,
\bea
\tilde\L_i~=~ -\p^j \Pi_{i j}~+~\fr{m}{2}\e_{0 i j k}F^{j k}~
+~\p_i p\we 0,\\
\tilde{\w}=\p_j B^{i j}~+~\p^i q \we 0. 
\eea
\nn Note that constraints are no longer reducible.
The primary first class constraints have the gauge -fixing conditions,
\bea
A_0\we0,\\
B_{oi}\we0.
\eea
The non-vanishing Dirac brackets, computed using the modified constraints
are now given by,
\be
\left\{ A_i(x)~,~ \Pi^l(y)\right\}^*= \left(\d_i^l~+~\fr{\p^l \p_i}
{\nabla^2}\right)\d(x-y),
\ee
\be
\left\{ B_{i j}(x)~,~\Pi^{lm}(y)\right\}^*=\left[\left(\d_i^l \d_j^m - \d_i^m\d_j^l
\right) + 
\fr{1}{\nabla^2} \left(\d_i^l\p_j\p^m -\d_i^m\p_j\p^l-\d_j^l\p_i\p^m+
\d_j^m\p_i\p^l\right)\right]\d(x-y),
\ee
\be
\left\{\Pi_i(x)~,~ \Pi_{l m}(y) \right\}^*=-\fr{m}{\nabla^2} 
\left(\e_{0 i l j}\p_m -\e_{0 i m j}\p_l\right)\p^j \d(x-y).
\ee
(16, 17, and 18) are in agreement with the Dirac brackets obtained by
other methods \cite{neto}.

Note that the Dirac brackets are independent of the new variables 
$p ~{\rm and}~q$.
Once the Dirac brackets are evaluvated, we can implement the constraints
strongly. Hence, on the constraint surface, $~\p^i\tilde\L_i = -\nabla^2
p =0~$. This sets the auxiliary variable $p=0$. In the same way we get $~q=0.~$
Thus the new variables introduced to handle reducibility vanish on the 
constraint surface and do not appear in the final Dirac brackets.

Next we study in the same manner two other equivalent formulations of 
$B{\wedge}F$ theory (differing only by a total divergence), where the
topological term is $H{\wedge}A$ or ($B{\wedge}F~-~H{\wedge}A$). We refer
them as $H{\wedge}A$ and symmetric $B{\wedge}F$ formulations respectively.
The reason for this study is that these different formulations map
naturaly to the different formulations of NTMGTs.

Instead of 
$+ {\frac{m}{4}} \e_{\m\n\r\l}B^{\m\n}F^{\r\l}~$ term in (1), now we consider
$~-\fr{m}{2\times 3!}\e_{\m\n\r\l}H^{\m\n\r}A^\l~$ ($H{\wedge}A$ formulation).
The primary constraints remain the same, but the momenta for the
unconstrained fields are now,
\bea
\Pi_i = -F_{0 i},\\
\Pi_{ij}=~H_{0 i j}~-~m\e_{0 i j k} A^{k}.
\eea
and the structure  of the secondary constraints are
changed to 
\bea
\W=\p_i\Pi^i~+~\fr{m}{2}\e_{0 i j k}\p^i B^{j k}\\
\L_i = \p^j\Pi_{i j}.
\eea
\nn The canonical Hamiltonian now reads,
\be
{\cal H}_c = -\fr{1}{2}\Pi_i\Pi^i +\fr{1}{4}\Pi_{ij}\Pi^{ij} 
+ \fr{1}{4}F^{ij}F_{ij} - \fr{1}{2\times 3!} H_{ijk}H^{ijk} 
+ \fr{m}{2}\e_{oijk}\Pi^{ij}A^k -\fr{m^2}{2}A_iA^i
\ee
Note again that the second constraint (22) is a reducible one, which can be 
handled in the
same way as in (12) and (13). The Dirac brackets $\{A_i,\Pi^j\}^* {\rm and}
\{B_{ij}, \Pi^{lm}\}^*$ remains the same as (16) and (17) while that
between the momenta now reads,
\be
\left\{\Pi_i~,~\Pi_{l m}\right\}~=~\fr{m}{\nabla^2} \e_{0 l m j}
\p_i\p^j\d(x-y).
\ee

Next we give the constraint structure of (1), where the 
mass term is written
symmetrically as $+ {\frac{m}{8}} \e_{\m\n\r\l}B^{\m\n}F^{\r\l}~+
~-\fr{m}{4\cdot 3!}\e_{\m\n\r\l}H^{\m\n\r}A^\l~$. Primary constraints
again remain same and the secondary constraints
are now,
\bea
\W =\p^i\Pi_i + \fr{m}{4}\e_{0ijk}\p^iB^{jk},\\
\L_i = \p^j\Pi_{ij} +\fr{m}{4}\e_{0ijk}F^{jk}.
\eea
\nn Here too, one of the constraints (26) is reducible.
The canonical Hamiltonian will now read
\bea
{\cal H}_c = -\fr{1}{2}(\Pi^i + \fr{m}{4}\e^{0ijk}B_{jk})
(\Pi_i + \fr{m}{4}\e_{oilm}B^{lm})
+\fr{1}{4}(\Pi_{ij} + m\e_{oijk}A^k)(\Pi^{ij}+m\e^{0ijl}A_l) \nonumber\\
+\fr{1}{4}F^{ij}F_{ij} - \fr{1}{2\times 3!} H^{ijk}H_{ijk}
\eea
The non-vansihing Dirac bracket (16) and (17) remains same while that between 
the momenta will now become
\be
\{\Pi_i, \Pi_{jk}\}^* = \fr{m}{2\nabla^2}\left[\e_{0jkl}\p_i\p^l
-(\e_{0jmi}\p_k -\e_{0kmi}\p_j)\p^m\right]\d({\vec x}-{\vec y})
\ee
\nn {\bf Section II}

In this section we quantise the three different St\"uckelberg formulations
(A, B, and C) 

The Lagrangian describing (A) is 
%abelian first-order St\"uckelberg
%formulation of massive spin-one theory is given by
\bea
{\CL} = -\fr{1}{4}(H_{\m\n}-C_{\m\n})(H^{\m\n}-C^{\m\n}) +\fr{1}{2}(G_\m +
\p_\m \T)(G^\m + \p^m \T) \nonumber\\
- \fr{1}{4m}\e_{\m\n\l\s}H^{\m\n}\p^\l G^\s -
\fr{1}{4m}\e_{\m\n\l\s}\p^m H^{\n\l} G^\s,
\eea
\nn where $C_{\m\n} = (\p_\m C_\n -\p_\n C_\m).$
\nn This Lagrangian is invariant under the transformations$
\d(G_\m)=\p_\m \L,~ \d(\T)=-\L,$ and
$~\d(H_{\m\n})=(\p_\m\L_\n -\p_\n\L_\m),~\d(C_\m) = \L_\m.~$
Note that when $\L_\m =\p_\m\w$, $H_{\m\n}$ is invariant implying the
reducible nature of the constraints.

The primary constraints following from this Lagrangian are

\bea
\Pi_0,~~\Pi_{0i},~~{{\t \Pi}_0},\\
\W_i = (\Pi_i - \fr{1}{4m}\e_{oijk}H^{jk}),\\
\L_{ij}= (\Pi_{ij} + \fr{1}{2m}\e_{0ijk}G^k).
\eea

The canonical Hamiltonian and the secondary constraints are
\be
{\cal H}_c= \fr{1}{4}(H_{ij}-C_{ij})(H^{ij}-C^{ij}) 
-\fr{1}{2}(G_i + \p_i\T)(G^i+\p^i\T)
-\fr{1}{2}{{\t \Pi}_i}{{\t \Pi}^i} +\fr{1}{2} (\Pi_\T)^2,
\ee
\bea
\L = \Pi_\T + \fr{1}{2m}\e_{0ijk}\p^i H^{jk},\\
\L_i = -{{\t \Pi}_i} + \fr{1}{m}\e_{0ijk}\p^jG^k,\\
\w = (\p^i{{\t \Pi}_i})
\eea
In the above, ${{\t \Pi}_\m}$ are the conjugate momenta corresponding to $C_\m$
fields and $\Pi_\T$ is the conjugate momentum of $\T$. Following Fadeev and 
Jackiw, \cite{fj} we imppose the symplectic 
\nn conditions(31) and (32) strongly. This results a non-vanishing bracket
\be
\{ G_i~,~ H_{ij} \}_{pb} = -m\e_{0ijk} \d({\vec x} - {\vec y}).
\ee
Since $\p^i\L_i + \w =0,$ this theory is reducible. 
%Here we see that the reducibility arises not because 
%of the linear
%dependence of the different components of the constraint $~\L_i~$
% that generates
%time independent transformation for the second rank field and compensating
%transformation for the coresponding St\"uckelberg field, but because
%$\L_i$ and the constraint that generates the independent trasformation for
%the vector St\"uckelberg field $~\w~$, are linearly dependent.

As in the previous section, we introduce auxiliary phase space variables
$p~{\rm and}~q$ obeying (11) and modify $~\L_i$ to $ {{\t \L}_i} +\p_i p~$ 
to handle reducibility.

We choose the gauge fixing conditions to be
\bea
\p^i G_i\we 0,\\
\p_l H^{ml} + \p^m q\we 0,\\
\p^iC_i\we 0.
\eea

Now the non-vanishing Dirac bracekts are
\be
\{G_i(x)~,~H_{jk}(y)\}^*=-m \left[\e_{oljk}(\d_i^l + \fr{\p_i\p^l}{\nabla^2})
+ \fr{1}{\nabla^2}(\e_{oijl}\p_k - \e_{0ikl}\p_j)\p^l\right] \d(\vec x - 
\vec y),
\ee
\be
\{G_i(x)~,~C_j(y)\}^* = \fr{m}{\nabla^2} \e_{oijk} \p^k \d(\vec x - \vec y),
\ee
\be
\{H_{ij}(x)~,~\th(y)\}^*=\fr{m}{\nabla^2}\e_{oijk}\p^k\d(\vec x - \vec y),
\ee
\be
\{ {\t \Pi}_i(x)~,~C^j(y)\}^* = -\left[ \d_i^j + \fr{\p_i \p^j}{\nabla^2}
\right] \d(\vec x - \vec y).
\ee

%\nn {\large Proca St\"uckelberg Formulation}

Next we consider the Lagrangian describing (B),
\be
{\CL}= -\fr{1}{4} F_{\m\n} F^{\m\n} + \fr{1}{2} (m A_\m -{\p_\m} \Phi)
(m A^\m- {{\p}^\m} \Phi).
\ee
This is the well studied St\"uckelberg - Proca Lagrangian and 
not a 
reducible
system. We give the Dirac brackets, which are rather well known.
The Hamiltonian follwoing from this Lagrangian is 
\be
{\cal H}_c = -\fr{1}{2}\Pi_i\Pi^i +\fr{1}{2}{\t \Pi}{\t \Pi} +\fr{1}{4}
F_{ij}F^{ij} -\fr{m^2}{2} A_iA^i +mA_i\p^i\Phi -\fr{1}{2}\p_i\Phi\p^i\Phi,
\ee
and the constraints $\Pi_0\we0,~ \w = (\p^i\Pi_i- m{\t\Pi})\we 0,$
are first class. The non-vanishing Dirac
\nn brackets evaluvated in Coloumb gauge are
\bea
\{A_i~,~ \Pi^j\}^* = ({\d^j_i} +\fr{\p_i\p^j}{\nabla^2}) \d({\vec x}-{\vec y}),\\
\{\t\Pi~,~ \Phi\}^* =-\d({\vec x}-{\vec y}),\\
\{\Pi_i~,~ \Phi\}^* = \fr{m}{\nabla^2} \p_i \d({\vec x}-{\vec y}).
\eea

%\nn {\large Kalb-Ramond St\"uckelberg Formulation}

The Lagrangian describing (C) is 
\be
{\CL} = \fr{1}{2\times3!} H_{\m\n\l}H^{\m\n\l} - \fr{1}{4}
(m B_{\m\n} - \Phi_{\m\n})(m B^{\m\n} - \Phi^{\m\n}),
\ee
\nn where $\Phi_{\m\n} = (\p_\m\Phi_\n -\p_\n \Phi_\m)$, and $\Phi_\m$ is
the St\"uckelberg field.
The canonical Hamiltonian following from this 
Lagrangian is
\be
{\cal H}_c=\fr{1}{4}\Pi_{ij}\Pi^{ij} -\fr{1}{2}\Pi_i\Pi^i -
\fr{1}{2\times3!}H_{ijk}H^{ijk} +\fr{1}{4}(mB_{ij} - \phi_{ij})
(mB^{ij} - \phi^{ij}),
\ee
\nn The persistence of the primary constraints$~\Pi_0\we0,~{\rm and}~
\Pi_{0i}\we0~$ give the Gauss law constraints,
\bea
\w=\p^i\Pi_i\we0,\\
\L_i=-(\p^j\pi_{ij} + m\pi_i)\we0.
\eea
All these constraints are first class. Here, we have $\p^i\L_i + m\w =0,$
which implies that $\L_i ~{\rm and}~ \w$ are not linearly independent and
thus the St\"uckelberg formulation given by the Lagrangian (50) describes a
reducible gauge system.

Here also we handle reducible constraints and their gauge fixing
conditions in the same way as we have done in the other two reducible theories
by enlarging the phase space by introducing
auxilliary variables $p ~{\rm and}~ q$ obeying (11).
The modified constraint and gauge fixing condition are
\bea
\t{\L}_i = -\p^j\Pi_{ij} -m\Pi_i +\p_i p,\\
\t {\T}_i = \p^lB_{il} +\p_iq,
\eea
\nn and corresponding to the generator $\w $, we choose the gauge condition
\be
\p^i \Phi_i\we0.
\ee

The non-vanishing Dirac brackets are
\be
\{\Pi_{ij}(x)~,~B^{lm}(y)\}^* = -\left[(\d^l_i\d^m_j -\d^m_i\d^l_j)
+\fr{1}{\nabla^2} (\p_j\p^m \d^l_i - \p_j \p^l \d^m_i - \p_i \p^m \d^l_j
+ \p_i \p^l \d^m_j)\right] \d(\vec x-\vec y),
\ee
\be
\{ \Pi_i (x)~, \Phi^j (y)\}^*= -(\d^j_i + \fr{\p_i
\p^j}{\nabla^2})\d(\vec x-\vec y),
\ee
\be
\{\Pi_{ij}(x)~,~\Phi^l(y)\}^*= \fr{m}{\nabla^2}(\p_i\d^l_j
-\p_j\d^l_i)\d(\vec x-\vec y).
\ee

This completes the Hamiltonian analysis of NTMGTs. 

%\newpage
\vskip0.5cm
\nn {\bf Section III}

In this section, from the considerations of the algebra of phase space
variables in the reduced phase space of the NTMGTs and TMGT, we arrive at
the following corespondance between the canonical variables.

The mappings between the phase space variables of symmetric $B{\wedge}F$
theory and NTMGT (A) is 

\bea
(\Pi^l +\fr{m}{4}\e^{0lmn}B_{mn})_{BF}\rla 
\fr{1}{2}\e^{0lmn}(H_{mn}-C_{mn})_{A},\\
(\Pi^{lm} + \fr{m}{2}\e^{0lmn}A_n)_{BF}\rla
\e^{0lmn}(G_n + \p_n\T)_{A},\\
(\fr{1}{3!} \e^{0ijk} H_{ijk})_{BF} \rla 
(\Pi_{\T})_{A},\\
(\fr{1}{2}\e_{0ijk}F^{jk})_{BF}\rla({\t{\Pi}_i})_{A},\\
(A_i)_{BF}\rla\fr{1}{m}(G_i)_{A},\\
(B_{ij})_{BF}\rla\fr{1}{m}(H_{ij})_{A}.
\eea
Under this map the Dirac brackets (41),(42),(43)and (44) and the
Hamiltonian (33) of the St\"uckelberg theory (29) gets mapped to that
of symmetric 
$B{\wedge}F$ theory.

In the case of the St\"uckelberg formulation described by(45), 
the mapping
\bea
(\Pi_{ij})_{BF}\rla (-\e_{oijk}\p^k\phi)_{B},\\
(B_{ij})_{BF}\rla(-\fr{1}{\nabla^2}\e_{oijk}\p^k {\t \Pi})_{B},\\
%(\fr{1}{2}\e_{0ijk}\p^iB^{jk})_{BF}\leftrightarrow(\t {\Pi})_{B},\\
%(\fr{1}{2\nabla^2}\e_{0ijk}\p^i \Pi^{jk})_{BF}\leftrightarrow(\Phi)_{B},\\
(A_i)_{BF}\leftrightarrow(A_i)_{B},\\
(\Pi_i)_{BF}\leftrightarrow(\Pi_i)_{B},
\eea
transform the Dirac brackets(47), (48) and (49) and the Hamiltonian of (46) 
to that  of $H{\wedge}A$ theory(23).

From the Dirac bracket structure of (50) and $B{\wedge}F$ the following 
mappings can be obtained,
\bea
(\Pi_i)_{BF} \leftrightarrow(\e_{0ijk}\p^j\Phi^k)_{C},\\
(A_i)_{BF}\leftrightarrow(\fr{1}{\nabla^2}\e_{0ijk}\p^j\Pi^k)_{C}  ,\\
(B_{ij})_{BF}\leftrightarrow(B_{ij})_{C},\\
(\Pi_{ij})_{BF}\leftrightarrow(\Pi_{ij})_{C},
\eea
\nn under which the Dirac brackets(57), (58) and (59) and the Hamiltonian of 
St\"uckelberg formulation(51) gets 
mapped to that of $B{\wedge}F$ theory (1).

Thus in the case of the mapping from
TMGT to (B), the two form field and its conjugate momentum undergo a
canonical transformation and in the case of mapping from TMGT to (C), the
vector field and its momentum transform canonically.

It can be checked that these identifications also map their
respective Hamiltonian equations of motions.

\nn {\bf Section IV}

We have found the mappings betwen the canonical variables of NTMGTs (A),
(B), and (C) to $B{\wedge}F$, $H{\wedge}A$ and symmetric
$B{\wedge}F$ formulations respectively. This is consistent with our
earlier observation that the $B{\wedge}F$ and $H{\wedge}A$ formulations go
over to (C) and (B) respectively by Buscher's duality procedure.
 This is understandable as Buscher's procedure is
itself is known to be a canonical transformation \cite{alv}. The first 
order St\"uckelberg theory 
was also shown to be equivalent to symmetric $B{\wedge}F$ theory by phase space
path integral approach, which is in agreement with what is observed here.
Thus this analysis in the canonical approach neatly complements the 
earlier studies.

The Gauss law constraint (7) and its gauge fixing condition in the case
of $B{\wedge}F$ theory can be
solved uniquely on the reduced phase space to get $\Pi_i =\e_{oijk}\p^j\psi^k$
and $A_i = \e_{0ijk}\p^j \chi^k$. Here $\psi_i$ is an
arbitrary field and $\chi_k$ is its conjugate momentum. These solutions, after
idetifying $\psi_i~{\rm and} \chi_i$ with the St\"uckelberg field and
its momentum (scaled by a factor of $\fr{1}{\nabla^2}$) are same as
the mappings (70) and (71). In the case of $H{\wedge}A$ theory
also we can solve the Gauss law constraint (22) and its gauge fixing
condition unquiely. These solutions, also provide 
the same mappings (66) and (67). Thus the mapping
of TMGTs to NTMGTs depend on the form of the Gauss
law constraints.

We can also bring out the identification between the fields of in TMGT and
NTMTs in a covariant way in Lagrangian formulation. For this purpose,
couple the gauge invariant (dual) field strengths ${\t F}_{\m\n} ~{\rm and}~
{\t H}_\m$ in $B{\wedge}F$ theory to sources $J^{\m\n} ~{\rm and}~ J^\m $
respectively.  By following the Buscher's duality procedure, as in \cite{us},
we get the dual Lagrangian to be that of (50), with additional source
dependent terms of the form $$ J^\m{\t {H_\m}}- \fr{1}{4} J_{\m\n}J^{\m\n} 
+\fr{1}{2} (mB_{\m\n} - \phi_{\m\n})J^{\m\n}.$$ The sources have to be
coupled to field strengths, rather than to the fields, so that there is a
global shift symmetry of the fields, needed for applying Buscher's procedure.
Thus we get the identification, 
at the level of correlation function
\be
\left< {\t F}_{\m\n}(x), {\t F}_{\l\s}(y)\right>_{BF}=
-\left<{\t B}_{\m\n}(x),{\t B}_{\l\s}(y)\right>_C + \left( g_{\m\l}g_{\n\s}-
g_{\m\s}g_{\n\l}\right)\d(x-y),
\ee
\nn where ${\t B}_{\m\n}=(mB_{\m\n}-\Phi_{\m\n}).$ Thus, apart from this 
contact term, we identify
${\t F}_{\m\n}$ in $B{\wedge}F$ theory to $(mB_{\m\n}-\phi_{\m\n})$ of
the St\"uckelberg formulation (C). Similar analysis can be done with the
other equivalent formulation of $B{\wedge}F$ theory (23), leading to the
identification
${\t{H_\m}}$ in $H{\wedge}A$ theory to $ (m A_\m -\p_\m\Phi)$ of the
St\"uckelberg formulation of Proca theory (B). Also by starting with the
first-order Lagrangian (29) and using the same procedure we identify
$(G_\m + \p_\m\T)$ and $(H_{\m\n}-C_{\m\n})$ of (29) to ${\t H}_\m$ and
${\t F}_{\m\n}$ respectively, of the symmetric $B{\wedge}F$ theory.
All these identification are modulo non-propagating contact terms. Note that 
in all cases
gauge invariant filed strengths of TMGT get mapped to gauge invariant
combination of fields in NTMGTs.

It will be interesting to see whether the
equivalence established here for free theories will hold good for 
the interacting as well as for the non-abelian generalisations of 
the models studied in this paper. Work along these lines is in progress.
\vskip0.5cm
\noindent {\bf Acknowledgements}:
We thank R. Banerjee for initial colloboration. We also thank
V. Srinivasan for encouragement. EH thanks 
U.G.C., India for support through S.R.F scheme.

\end{document}